\documentclass[10pt,twocolumn,showpacs,longbibliography,amsmath,amssymb,prd,floatfix,superscriptaddress]{revtex4-2}

\usepackage{graphicx}
\usepackage{dcolumn}
\usepackage{bm}
\usepackage{color}
\usepackage{txfonts}
\usepackage{microtype}
\usepackage{epstopdf}
\usepackage{pict2e}     
\newcommand\numberthis{\addtocounter{equation}{1}\tag{\theequation}}
\usepackage{tikz-feynman} 
\usepackage{subfigure} 
\definecolor{plotblue}{RGB}{31,119,180}  
\definecolor{plotred}{RGB}{204,85,34}
\begin{document}

\title{Spin-Dependent Axion Generation with Controllable Emission Angles in Strong Laser Fields}

\author{Jia-Ding Chen}
\affiliation{Department of Physics, Shanghai Normal University, Shanghai 200234, China}
\author{Hang Liu}
\affiliation{Department of Physics, Shanghai Normal University, Shanghai 200234, China}
\author{Kai-Hong Zhuang}
\affiliation{Department of Physics, Shanghai Normal University, Shanghai 200234, China}
\author{Baifei Shen}
\affiliation{Department of Physics, Shanghai Normal University, Shanghai 200234, China}
\author{Pei-Lun He}
\email{peilunhe@sjtu.edu.cn}
\affiliation{State Key Laboratory of Dark Matter Physics, Key Laboratory for Laser Plasmas (Ministry of Education) and School of Physics and Astronomy, Collaborative Innovation Center for IFSA (CICIFSA), Shanghai Jiao Tong University, Shanghai 200240, China}
\author{Yue-Yue Chen}
\email{yueyuechen@shnu.edu.cn}
\affiliation{Department of Physics, Shanghai Normal University, Shanghai 200234, China}
\date {\today}

\begin{abstract}
We investigate axion production in the collision between a spin-polarized relativistic electron beam and an ultraintense laser pulse. A spin-resolved Monte Carlo framework is developed to model axion–electron and axion–photon couplings in arbitrary electromagnetic fields, using quantum emission probabilities under the local constant field approximation. Owing to spin-dependent asymmetries in radiation probability, the emitted axions acquire a characteristic angular deflection tied to the initial electron polarization. This spin-dependent asymmetry enables control over the axion emission direction by adjusting the polarization of the electron beam and laser field. Simulations show that a dense and collimated axion beam ($\sim 10^{10} g_{ae}^2$) with a tunable deflection angle ($\sim$ mrad) can be produced within tens of femtoseconds using current laser technology. Our results establish a novel mechanism for manipulating axion trajectories and open a promising route toward laboratory-based searches for the axion–electron coupling.
\end{abstract}

\maketitle

The identity of dark matter remains one of the central questions in contemporary science \cite{kolb1990early}. One of the leading candidates is the QCD axion \cite{preskill1983cosmology,abbott1983cosmological,dine1983not}, a hypothetical particle first proposed by Weinberg and  Wilczek  to solve the strong CP problem in the Standard Model of particle physics \cite{weinberg1978new,wilczek1978problem,peccei1977cp,peccei1977constraints}. In addition to the QCD axion, similar particles called axion-like particles (ALPs) also emerge in string theory \cite{witten1984some,arvanitaki2010string} and supersymmetric extensions \cite{baer2019gravity}. 
Various experimental searches have been conducted and are ongoing to detect ALPs and impose stricter limits on ALP models. These limits can be broadly categorized into three types: cosmological \cite{sikivie1994axion,caldwell2017dielectric,graham2018spin,
ouellet2019first}, astrophysical \cite{sikivie1983experimental,sikivie1985detection,
avignone1998experimental,akerib2017first}, and laboratory-based \cite{ballou2015new,bahre2013any,srednicki2012particle,della2016pvlas,konaka1986search,davier1989unambiguous,bross1991search,abrahamyan2011search,dobrich2016alptraum,kim2010axions,andreas2012new,baker2013quest,irastorza2018new}. While cosmological and astrophysical limits generally provide stronger constraints than laboratory-based searches, they are highly model-dependent and can be significantly weakened 
 \cite{masso2006compatibility,jaeckel2007need}. 
The vulnerability in the astro-cosmological constraints  provides a strong motivation to pursue well-controlled laboratory-based searches.


Laboratory-based searches for axion-like particles include light-shining-through-walls (LSW) experiments such as OSQAR \cite{ballou2015new} and ALPS \cite{bahre2013any}, which rely on photon-axion conversion and subsequent regeneration across an opaque barrier \cite{srednicki2012particle}. The PVLAS collaboration \cite{della2016pvlas} probes axion-induced vacuum dichroism by measuring the polarization rotation of a laser beam propagating through a magnetic field. However, the sensitivity of these setups is fundamentally limited by the achievable magnetic field strength ($|B| \sim 10^5$ G), the finite interaction length ($L \sim 1$ km), and the requirement of phase coherence between the photon and axion waves. Furthermore, in PVLAS, the tree-level axion signal is superimposed with the QED-induced vacuum birefringence, making signal isolation particularly challenging \cite{melnikov2021dispersion}. These limitations underscore the need for alternative approaches that can overcome coherence constraints, suppress QED backgrounds, and access unexplored regions of the axion parameter space.

The advent of ultraintense laser systems enables electromagnetic field strengths with 
\( B\sim10^{10} \mathrm{G} \), exceeding those of conventional static-field setups by orders of magnitude~\cite{ELI,XCELS}. These extreme conditions, together with high photon fluxes and short temporal scales, offer new opportunities to explore ALP interactions in previously inaccessible regimes. In particular, the large momentum transfer in laser--beam collisions extends sensitivity to higher ALP masses~\cite{mendoncca2007axion,gies2009strong,dobrich2010axion,dobrich2010high,bai2024coherent}.
To date, theoretical investigations have predominantly focused on ALP--photon couplings~\cite{beyer2022light,beyer2020axion,villalba2016light,villalba2013axion}. However, several well-motivated extensions of the Standard Model predict direct couplings to electrons-such as the Arion~\cite{anselm1982second}, DFSZ-like scenarios~\cite{dias2014quest}, and string-inspired ALPs~\cite{cicoli2012type}-which remain largely unexplored. These couplings, in principle, allow for direct ALP production through collisions between relativistic electrons and intense laser pulses, opening a complementary channel for laboratory-based ALP searches.

The potential of intense laser pulses for probing electron-ALP coupling has been investigated in the low-energy and coherent limit. It shows that the degradation of the exclusion bound on $g_{ae}g_{a\gamma\gamma} (\text{GeV}^{-1})$ can be push to 10–100 eV, using nonlinear Thomson scattering of light scalars in a monochromatic electromagnetic background \cite{dillon2019light}. 
Moreover, the electron-seeded ALP production and subsequent decay of ALPs  in a monochromatic, circularly-polarised electromagnetic background are investigated \cite{king2018electron,ma2024laser,bai2022new}. However, the probabilities are presented as a summation over the photon number $n$, whose evaluation represents a tremendous task at high-field limit ($a_0\gg1$), limiting its application in strong-field simulations. A solution to this problem is adopting local constant field approximation (LCFA) \cite{ritus1985quantum}, that is commonly used for strong-field QED studies. The radiation probability of ALPs has been obtained, however, after averaging  the initial spin and summing over final spin of electrons \cite{dillon2018alp}. Recent studies have demonstrated that the spin effects during strong-field QED process have significant consequences on final particles polarization, and electron kinetic motion \cite{dai2024fermionic,elkina2011qed,ridgers2014modelling,green2015simla,gonoskov2015extended,chen2019polarized,li2020production,dai2022photon,zhuang2023laser,song2022dense,gong2023electron,xue2023generation,chen2025angular}. 
The spin degree of freedom—though fundamental in strong-field QED—has not yet been leveraged in axion production models, leaving a promising direction largely unexplored.

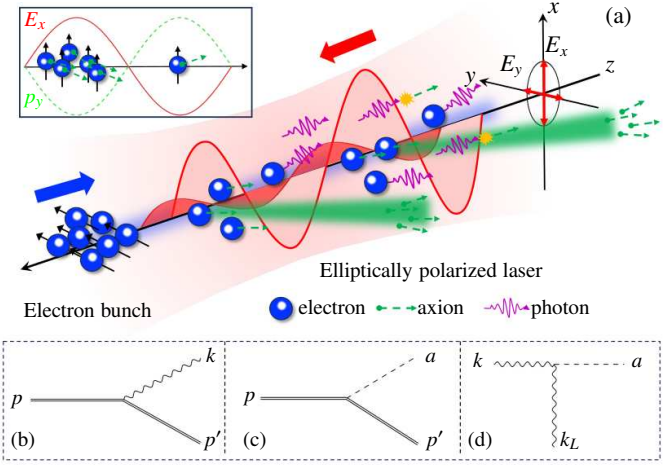
\begin{figure}[t]
    \centering
    \includegraphics[width=0.5\textwidth]{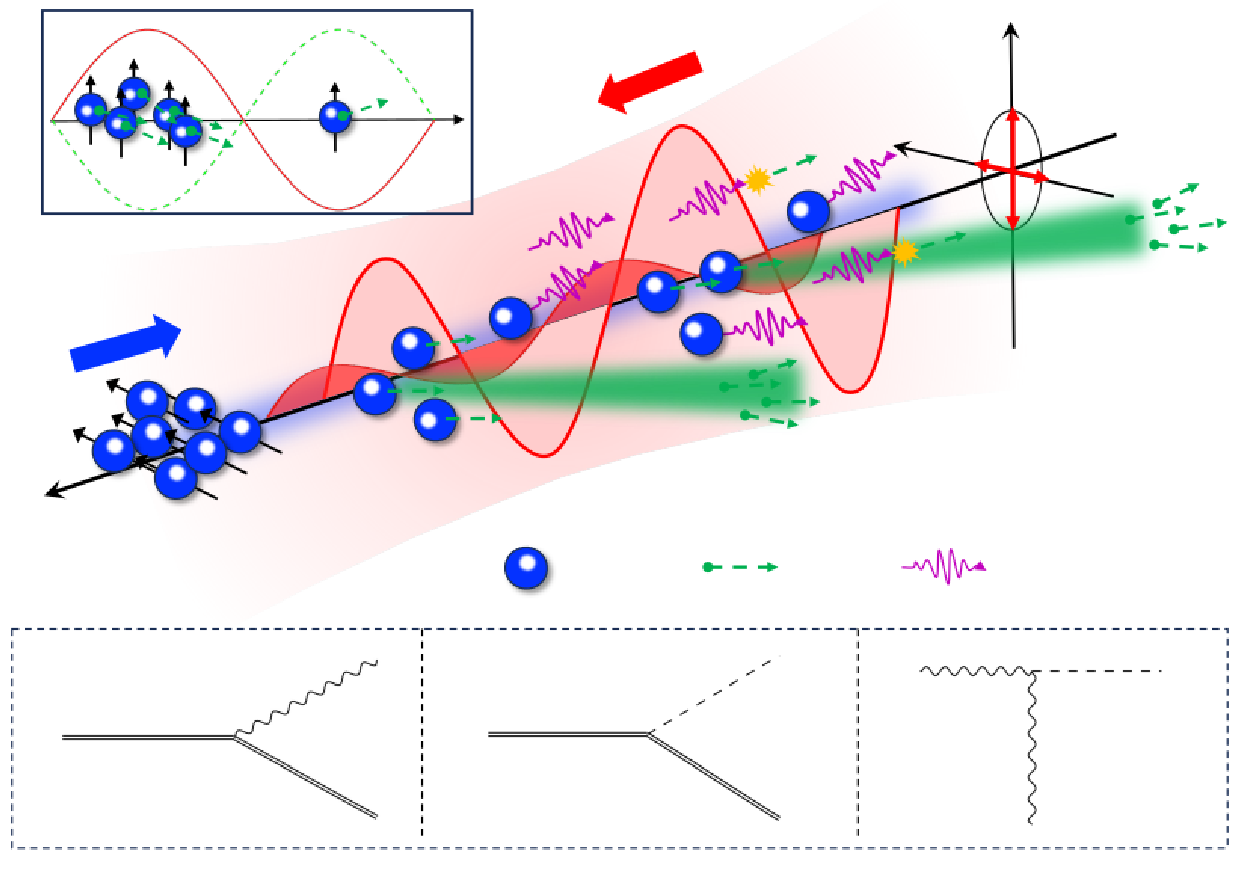} 
     \begin{picture}(300,17)
    \put(10,187){\footnotesize {\color{red}$E_x$}}
    \put(10,158){\footnotesize {\color{green}$p_y$}}
    \put(10,77){\footnotesize Electron bunch}
    \put(122,92){\footnotesize Elliptically polarized laser}
    \put(202,78){\footnotesize photon}
    \put(159,78){\footnotesize axion}
    \put(114,78){\footnotesize electron}
    \put(230,188){(a)}
    \put(230,170){$z$}
    \put(208,192){$x$ }
    \put(177,166){$y$}
    \put(190,170){$E_y$}
    \put(207,177){$E_x$}
    \put(5,28){\footnotesize (b)}
    \put(5,44){\footnotesize $p$}
    \put(79,60){\footnotesize $k$}
    \put(78,28){\footnotesize $p'$}
    \put(93,28){\footnotesize (c)}
    \put(93,45){\footnotesize $p$}
    \put(162,28){\footnotesize $p'$}
    \put(162,60){\footnotesize $a$}
     \put(178,28){\footnotesize (d)}
    \put(180,58){\footnotesize $k$}
    \put(240,58){\footnotesize $a$}
    \put(213,28){\footnotesize $k_L$}
    \end{picture}
    \caption{ (a) Illustration of axion generation schemes in strong laser fields. The electrons with initial polarization of $\zeta_y=1$ collide with an elliptically polarized laser pulse. The insert demonstrates that the axions are selectively generated at the positive half cycle ($E_x>0$) and emitted towards $p_y<0$.  Feynman diagrams illustrating photon emission (b) and axion production via electron-axion coupling (c) and photon-photon coupling (d).}
    \label{fig. scheme} 
\end{figure}

In this work, we studied the production of axions via nonlinear Compton scattering from polarized electrons in intense laser pulses (hereafter ``axions" refers to ALPs for brevity); see Fig. \ref{fig. scheme}. To this end, we have developed a spin-resolved Monte Carlo code that self-consistently incorporates the couplings of axions to electrons and photons, respectively. We find that the spin effects ignored in previous studies could induce a notable deflection of produced axions, which can be controlled by tuning laser ellipticity and electron spin polarization. The asymmetry in the angular distribution could serve as a distinctive signal of the axion-electron coupling [Fig. \ref{fig. scheme} (c)], clearly separated from its QED counterpart, i.e. photon emission [Fig. \ref{fig. scheme} (b)], as well as from dark counts.  In addition, the high single-shot yield and femtosecond-scale emission duration jointly contribute to an enhanced signal-to-noise ratio. Our scheme utilizes the polarization freedom to advance the control over laboratory-based searches of axion, and may provide  useful complementary bounds to those obtained in other lab-based and astrophysical experiments. 

In our simulations, axion production from electrons is calculated by integrating the spin-resolved emission probability [Eq.~(\ref{electron_axion})] along the electron trajectory, while axion generation from radiated photons is obtained by integrating the photon-axion conversion probability~\cite{SM3} along the photon path.
The electron trajectory is determined by classical dynamics governed by the Lorentz force and the Bargmann–Michel–Telegdi (BMT) equation, while incorporating quantum effects such as stochastic photon emission, spin polarization due to radiative spin-flip processes, and radiative corrections from one-loop electron self-energy \cite{SM1}.
The influence of axion emission on electron dynamics is statistically negligible due to the smallness of the coupling constant, \( g_{ae}\ll \alpha \sim \mathcal{O}(10^{-2}) \) \cite{SM2}.
The spin-dependent axion emission probabilities in LCFA are employed with the leading order contribution with respect to 1/$\gamma$, which are derived with the quantum electrodynamics operator method of Baier-Katkov. The probability of emitting an axion with energy of $\omega_a$
reads \cite{he2025semiclassical}
\begin{align*}\label{electron_axion}
\frac{dW_{a}}{d\omega_{a}dt}&=C_{a}\left\{ -\left[\frac{\varepsilon\varepsilon'}{2\omega_{a}^{2}}\delta_{m}^{2}\right]\textrm{IntK}_{\frac{1}{3}}\left(z_{q}^{a}\right)+\left(1+\frac{\varepsilon\varepsilon'}{\omega_{a}^{2}}\delta_{m}^{2}\right)\textrm{K}_{\frac{2}{3}}\left(z_{q}^{a}\right)\right.\\&\left.+\left(1+\frac{\varepsilon\varepsilon'}{\omega_{a}^{2}}\delta_{m}^{2}\right)^{\frac{1}{2}}\left(\vec{\zeta}\cdot\vec{b}\right)\textrm{K}_{\frac{1}{3}}\left(z_{q}^{a}\right)\right\}  .\numberthis
\end{align*}
Here, \( C_a = \frac{\sqrt{3}}{24\pi^{2}} \frac{g_{ae}^{2} \omega_a^{2} m_e^{2}}{\varepsilon^{3} \varepsilon'} \), \( z_{q}^{a} = \frac{2}{3} \frac{\omega_a}{\chi_{e} \varepsilon'} \left( 1 + \frac{\varepsilon \varepsilon'}{\omega_a^{2}} \delta_{m}^{2} \right)^{\frac{3}{2}} \), and \( \delta_{m} = \frac{m_{a}}{m_{e}} \), where \( m_a \) and \( m_e \) denote the masses of the axion and the electron, respectively. The energies of the electron before and after axion emission are denoted as \( \varepsilon \) and \( \varepsilon' \), respectively.  \( \vec{\zeta} \) is the  polarization vector of electron before emission.  The unit vectors \( \vec{v} \) and \( \vec{s} \) correspond to the velocity and acceleration directions, respectively, with \( \vec{b} = \vec{v} \times \vec{s} \). The integral of the modified Bessel function of the second kind is defined as \( \textrm{IntK}_{\frac{1}{3}}(z_q^a) \equiv \int_{z_q^a}^{\infty} dz \, \textrm{K}_{\frac{1}{3}}(z) \), where \( \textrm{K}_n \) represents the \( n \)-th order modified Bessel function of the second kind. $\chi_e$ is the nonlinear quantum parameter that controls the magnitude of quantum.
Averaging by the electron initial spin, the spin-unresolved radiation probability derived in previous literatures is obtained \cite{dillon2018alp}.

The simulation was performed with the following laser parameters: a peak intensity of \(I_0 \approx 2.16 \times 10^{22} \, \text{W/cm}^2\) (\(a_0 = 100\)), a wavelength of \(\lambda_0 = 800 \, \text{nm}\), a pulse duration of \(\tau = 10T_0\) (where \(T_0\) is the laser period), a focal radius of \(w_0 = 4 \, \mu\text{m}\), and an ellipticity of \(\epsilon = |E_y| / |E_x| = 0.05\). The cylindrical electron bunch has an angular divergence of \(0.3 \, \text{mrad}\) and an initial kinetic energy of \(\varepsilon_0 = 2 \, \text{GeV}\) with a relative energy spread of \(\Delta \varepsilon_0 / \varepsilon_0 = 0.05\).  The maximum quantum parameter reaches \(\chi_{\text{max}} \approx 1.5\). The bunch has \( N_e = 1 \times 10^6 \) electrons,  radius \(w_{e}=\lambda_{0}\), length \(L_{e}=5\lambda_{0}\), with a transversely Gaussian and longitudinally uniform distribution, which can be obtained by current laser wakefield accelerators. 

\begin{figure}[htbp]
 \centering
    \includegraphics[width=0.5\textwidth]{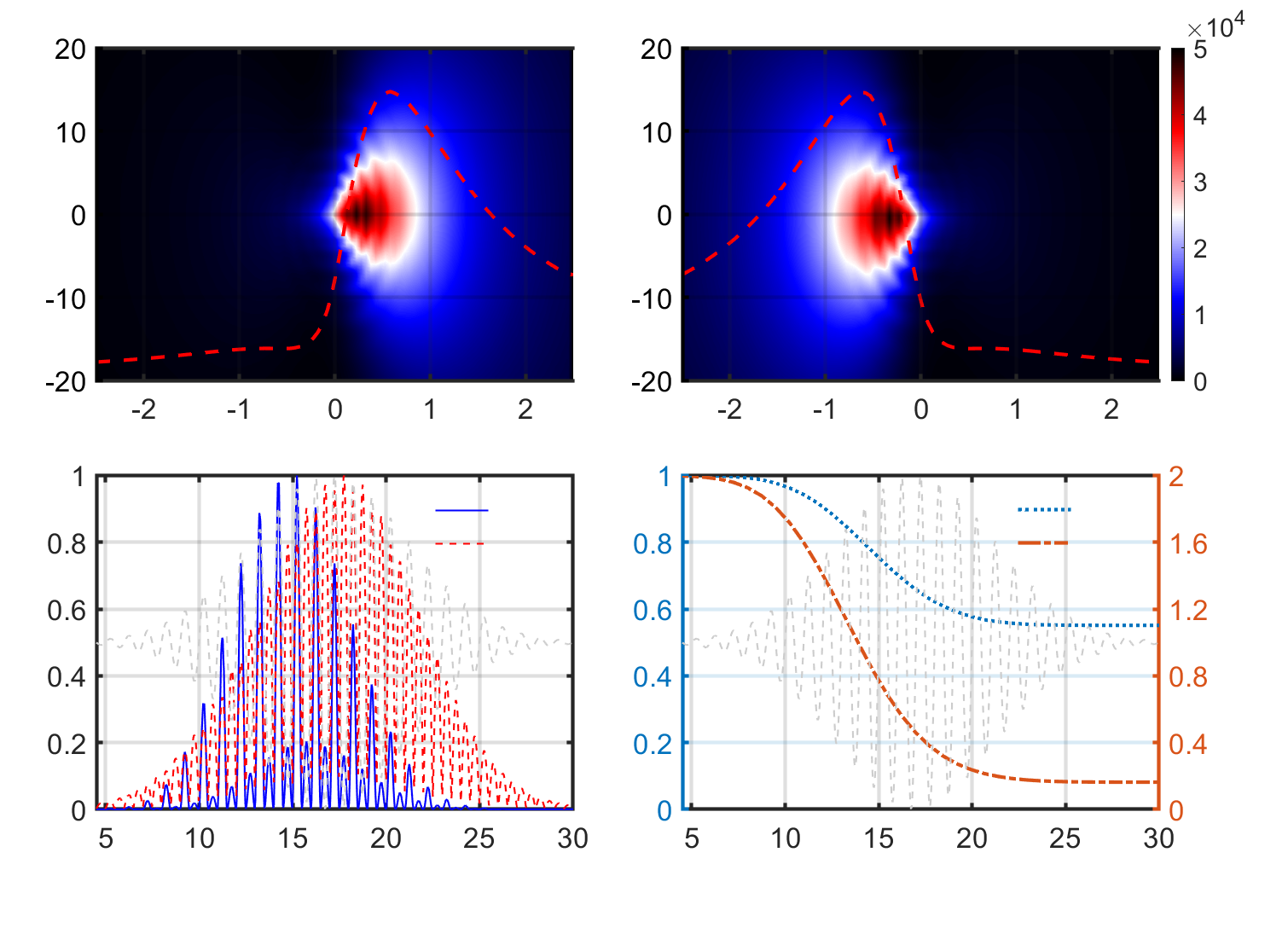} 
 \begin{picture}(300,20)
    \put(20,190){\textcolor{white}{(a)}}
    \put(137,190){\textcolor{white}{(b)}}
    \put(20,105){(c)}
    \put(137,105){(d)}
   \put(77.5,191){\color{white}{\circle*{3} }}
    \put(85,190){\color{white}{$\theta_{y,max}$}}
       \put(171,191){\color{white}{\circle*{3} }}
    \put(177.5,190){\color{white}{$\theta_{y,max}$}}
    
    \put(0,150){\rotatebox{90}{$\theta_x$ (mrad)}}
    \put(50,120){$\theta_y$(mrad)}
    \put(165,120){$\theta_y$(mrad)}
    \put(50,33){$\varphi/\left(2\pi\right)$}
    \put(98,107){\tiny{$N_a$}}
    \put(98,100){\tiny{$N_\gamma$}}
    \put(115,80){\rotatebox{90}{$\zeta_y$}}
    \put(240,100){\rotatebox{270}{$\overline{\varepsilon}$ (GeV)}} 
    \put(170,33){$\varphi/\left(2\pi\right)$}
    \put(215,107){\tiny{$\zeta_y$}}
    \put(215,100){\tiny{$\overline{\varepsilon}$}}
    \put(-3,50){ \small \rotatebox{90}{Normalized yield}}
    \end{picture}
    \caption{The angular distribution of axion density \( d^2(N/g_{ae}^2)/d\theta_x d\theta_y (\text{mrad}^{-2})\) for initial electron spin \( \zeta_y = 1 \) (a) and \( \zeta_y = -1 \) (b), plotted against the deflection angles \( \theta_x = \arctan k_{ax}/k_{az} \) and \( \theta_y = \arctan k_{ay}/k_{az} \). The red dashed lines indicate the angular distribution of axion density \( d(N/g^2_{ae})/d\theta_y \) versus \( \theta_y \) for \( \zeta_y = 1 \) [in (a)] and \( \zeta_y = -1 \) [in (b)].
    (c) Normalized axion number \( N_a \) (blue solid line) and $\gamma$ photon number \( N_\gamma \) (red dashed line) at production points, versus laser phase \( \varphi /(2\pi) \). (d) Left \( y \)-axis: polarization \( \zeta_y \) (blue dotted line) versus laser phase \( \varphi /(2\pi) \); Right \( y \)-axis: Electron average energy $\overline{\varepsilon}$ (red dashed-dot line)  versus laser phase \( \varphi /(2\pi) \). \( m_a = 0 \) . The superimposed gray lines in (c) and (d) represent the $x$ component of the electric fields. }
    \label{fig. agl} 
\end{figure}
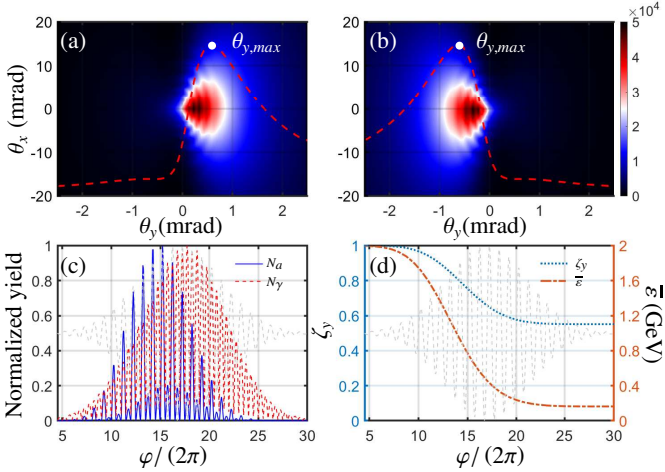

The initial spin polarization of electrons significantly influences the angular distribution of axions, as shown in Fig. \ref{fig. agl}. For  $m_a=0$, electrons with \(\zeta_y = 1\) predominantly generate axions with \(p_y^f < 0\) [Fig. \ref{fig. agl}(a)], while those with \(\zeta_y = -1\) produce axions with \(p_y^f > 0\) [Fig. \ref{fig. agl}(b)].
For electrons with an initial spin of \( \zeta_y = 1 \), the majority of axions are deflected to \( \theta_y > 0 \), accounting for 74.83\% of the axions generated within the range \( |\theta_x| < 20 \) mrad and \( |\theta_y| < 1 \) mrad. This behavior arises because the radiation probability for axions depends on the initial spin of the electron, described by:  
\[
w\left(\vec{\zeta}\right) = C_a \left[\text{K}_{\frac{2}{3}}(z_q^a) + (\vec{\zeta} \cdot \vec{b}) \text{K}_{\frac{1}{3}}(z_q^a)\right], \numberthis
\]  
where \(\vec{b}\) is a unit vector aligned with the local magnetic field direction in the considered setup.  
For electrons with \(\zeta_y > 0\), the radiation probability during the half-cycle when \(B_y > 0\) is significantly higher than during the half-cycle when \(B_y < 0\), resulting in a greater axion yield when \(B_y > 0\). The final momentum of the axion is determined by the electron momentum at the point of emission, i.e. $\vec{k}_{a}^f=\vec{p}\left(t_r\right)$. In an elliptically polarized laser field, the \(y\)-component of the electron momentum $p_y$ exhibits a \(\pi\)-phase delay relative to the magnetic field \(B_y \). Consequently, axions predominantly generated during \(B_y > 0\) acquire negative transverse momentum \(p_y\), corresponding to a positive emission angle \(\theta_y = \arctan(p_y / p_z)\) [Fig. \ref{fig. agl} (a)]. Conversely, for electrons with \(\zeta_y < 0\), the emission angle becomes \(\theta_y < 0\) [Fig. \ref{fig. agl} (b)]. The central deflection angel is approximately \( \theta_{y,max} \approx 0.5 \) mrad. Compared to $\gamma$ photon emission, axion production exhibits a much more pronounced cycle-dependent asymmetry [Fig. \ref{fig. agl} (c)], leading to a significantly enhanced angular asymmetry in axion-electron coupling compared to its QED counterpart, photon-electron coupling [Fig. \ref{Fig.garr} (a)].
Meanwhile, as electrons propagate through the laser pulse, they undergo substantial energy loss due to radiation reaction and depolarization caused by spin-flip transitions induced by photon emission [Fig. \ref{fig. agl} (d)]. Consequently, the cycle-dependent asymmetry gradually diminishes with increasing laser phase and eventually vanishes at the pulse tail [Fig. \ref{fig. agl} (c)]. This indicates that the asymmetry in the axion angular distribution is primarily governed by axion production in the early phase of the laser pulse, where electrons are highly energetic and polarized. 


The axion yield scales with the coupling parameter $g_{ae}$ and electron number $N_e$ as \( N_a \approx 3.03 N_e g_{ae}^2 \), indicating that the axion yield can reach \( N_a = 3.03 \times 10^{-10} \) for \( N_e = 1 \times 10^{10} \) and \( g_{ae} = 10^{-10} \). The single-shot axion flux can be estimated as \( \Psi = \frac{N_a}{\pi w_e^2 L_e} \approx 3.02 \times 10^{33} g_{ae}^2 \, \text{cm}^{-2} \, \text{s}^{-1} \)\cite{freidberg2008plasma} . For comparison, ALPs emitted by the Sun due to the Primakoff process are expected to have a flux density of \( \Psi = \left(\frac{g_{a\gamma\gamma}}{\text{GeV}^{-1}}\right)^2 \times 3.75 \times 10^{31} \, \text{cm}^{-2} \, \text{s}^{-1} \) at Earth \cite{andriamonje2007improved}. Therefore, a high-power laser facility is capable of generating axions with a comparable flux  in a single shot.

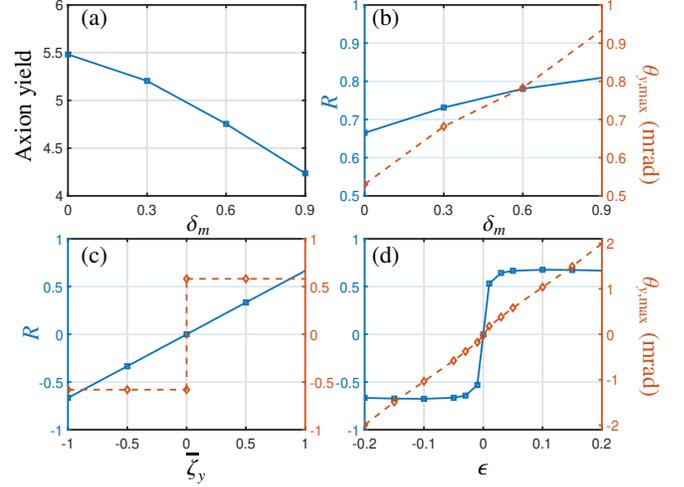
\begin{figure}
\hspace{-0.5cm} 
    \includegraphics[width=0.5\textwidth]{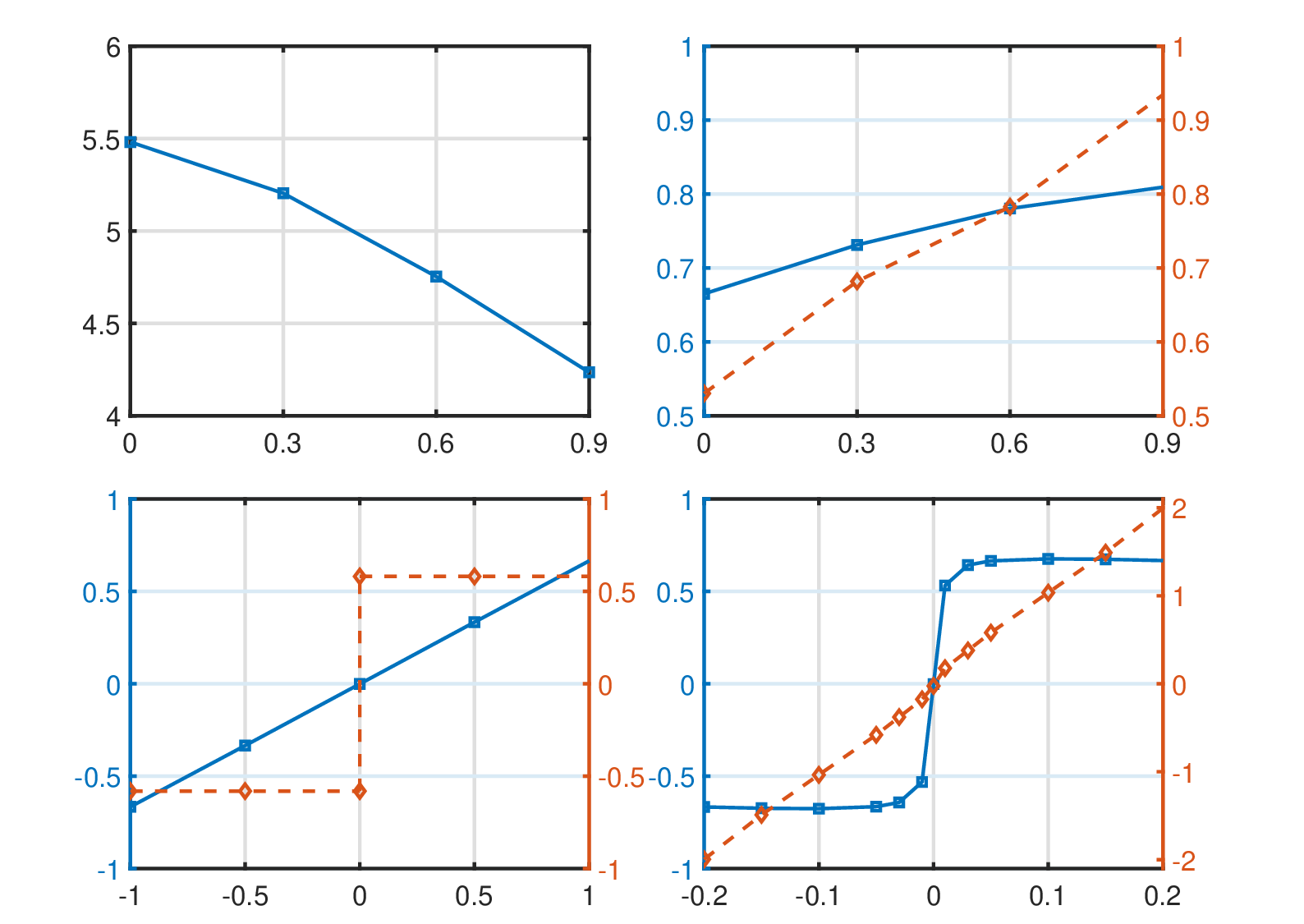}
     \begin{picture}(300,20)
    \put(20,185){(a)}
    \put(130,185){(b)}
    \put(20,95){(c)}
    \put(130,95){(d)}
    \put(-5,133){\rotatebox{90}{Axion yield }}
    \put(60,107){$\delta_m$}
    \put(111,153){\rotatebox{90}{{\color{plotblue}$R$}}}
    \put(230,170){\rotatebox{270}{\color{plotred}$\theta_{y,\max}$ (mrad)}}
    \put(172,107){$\delta_m$}
    \put(-2,66){\rotatebox{90}{{\color{plotblue}$R$}}}
    \put(60,15){$\overline{\zeta}_{y}$}
    \put(230,90){\rotatebox{270}{\color{plotred}$\theta_{y,\max}$ (mrad)}}
    \put(170,15){$\epsilon$}

    \end{picture}
\caption{ (a) The scaling law of axion yield log$_{10}\left(N_{a}/g_{ae}^2\right)$ versus $\delta_m$. The angular asymmetry $R$ (blue-solid line) and  deflection angle $\theta_{y,max}$ (red-dashed line)  versus: (b) $\delta_m$, (c) average polarization of electron $\overline{\zeta}_y$, and (d) ellipticity of laser $\epsilon$. The rest parameters are $\epsilon=0.05$, $\overline{\zeta}_y=1$ and $\delta_m=0$.  }
    \label{fig. gae}
\end{figure}

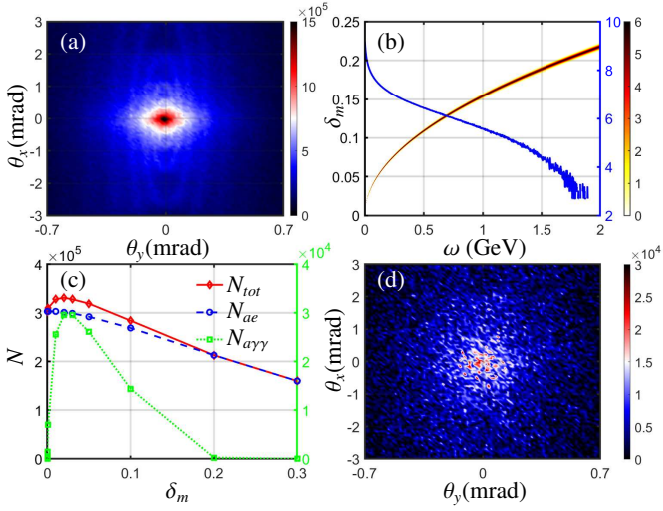
\begin{figure}[t]
    \includegraphics[width=0.5\textwidth]{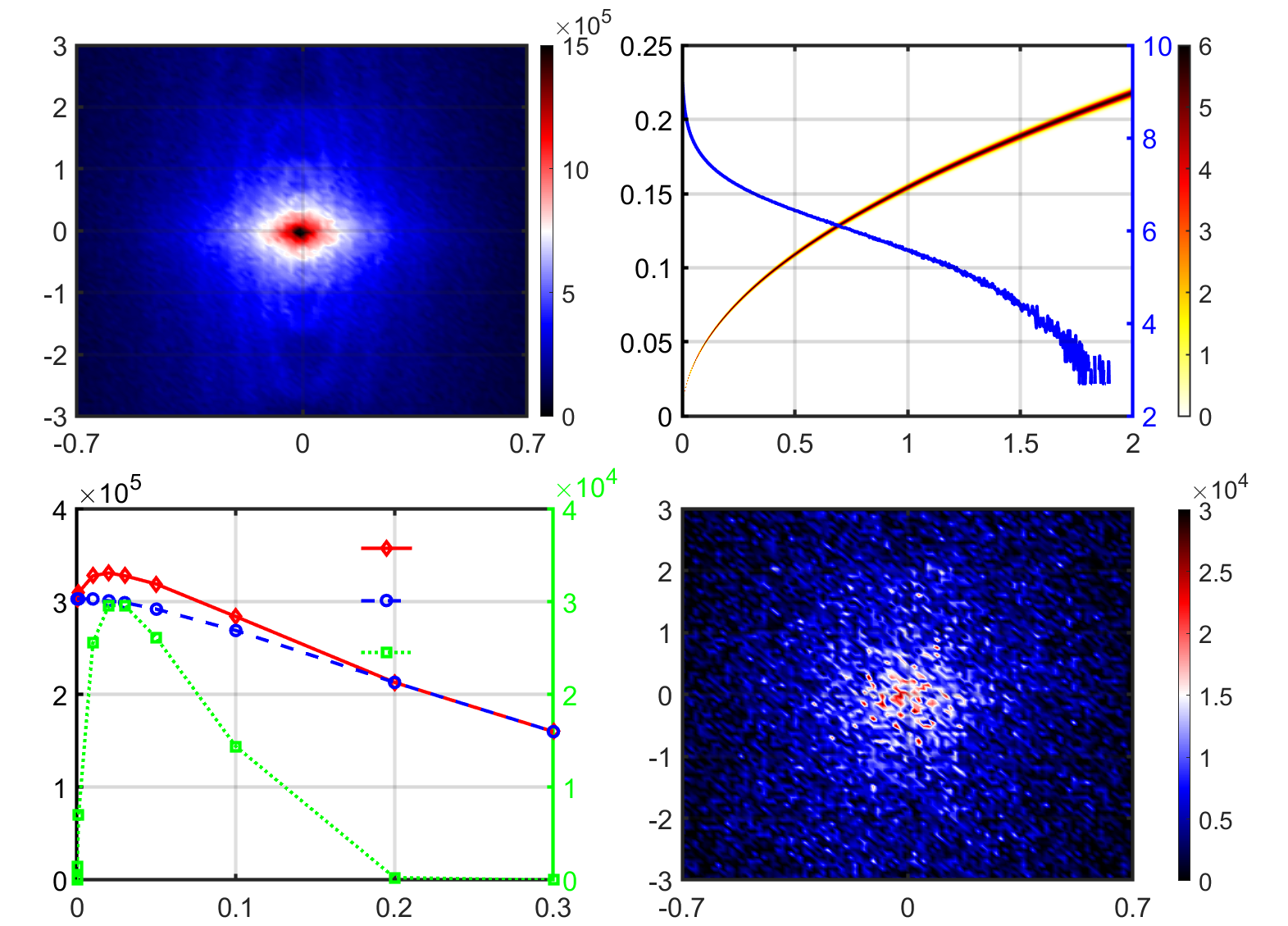}
     \begin{picture}(300,20)
    
    \put(20,185){\textcolor{white}{(a)}}
    \put(140,185){{(b)}}
    \put(20,95){(c)}
    \put(140,95){\textcolor{white}{(d)}}
    
    \put(0,140){\rotatebox{90}{$\theta_x$(mrad)}}
    \put(45,107){$\theta_y$(mrad)}

    \put(120,155){\rotatebox{90}{$\delta_m$}}
    \put(165,107){$\omega$ (GeV)}
    \put(0,65){\rotatebox{90}{$N$}}
    \put(60,15){$\delta_{m}$}

    \put(83,93){$N_{tot}$}
    \put(83,83){$N_{ae}$}
    \put(83,73){$N_{a\gamma\gamma}$}
    
    \put(120,55){\rotatebox{90}{$\theta_x$(mrad)}}
    \put(165,15){$\theta_y$(mrad)}

    \end{picture}
        \caption{(a) The $\gamma$ photons density $d^2 N_\gamma/d\theta_x d\theta_y$ (mrad$^{-2}$) versus $\theta_x$  and $\theta_y$. (b)  The $\gamma$ photons spectrum $\text{log}_{10} (dN_\gamma/d\omega)$ versus photon energy $\omega$ (blue solid line) and axion production probability 
       \( \text{log}_{10} (dW_{a\gamma\gamma}/g_{a\gamma\gamma}^2)/dt \)
      versus axion mass $m_a$ and $\gamma$ photon energy $\omega$. (c) Left $y$-axis: The total axion number $N_{tot}$ (red solid line) and the axion number generated via electron-axion coupling (blue dashed line) versus axion mass $\delta_m$;  Right $y$-axis: The axion yield via photon-axion coupling  (green dotted-dashed line)  versus axion mass $\delta_m$. (d) The axion density $d^2(N_a/g_{a\gamma\gamma}^2)/d\theta_x d\theta_y$ (mrad$^{-2}$) generated via Primakoff effect versus $\theta_x$  and $\theta_y$, for $\delta_m=0.03$. The rest parameters are same with Fig. \ref{fig. agl}.  }
        \label{Fig.garr}
\end{figure}

Meanwhile, the axion yield decreases with increasing $m_a$, owing to the suppression from the Bessel function in the radiation probability [Eq. (\ref{electron_axion})]. Specifically, the axion yield decreases by one order of magnitude as \( \delta_m \) increases from 0 to 0.9 [Fig. \ref{fig. gae} (a)]. Nevertheless, the spin-related asymmetry remains evident in the angular distribution, which can be defined as $R=\left(N_+-N_-\right)/\left(N_++N_-\right)$ with $N_\pm$ being the axion number located at $\theta_y>0$ and $\theta_y<0$, respectively. As $\delta_m$ increases, the angular asymmetry $R$ increases from 0.67 to 0.8, as well as the deflection angle \( \theta_y \) [Fig. \ref{fig. gae} (b)]. 
This occurs as the suppressed radiation probability for heavier axions confines their emission to regions of highest field strength—i.e., the pulse peak—where \( p_y \)  simultaneously attains its maximum.
In contrast, low-mass axions are produced across a wider laser phase interval, resulting in a smaller deflection angle $\theta_y$. As a consequence, the deflection angle for high-mass axions is greater than that for low-mass axions \cite{SM5}.

More interestingly, our scheme enables axion generation with a tunable angular asymmetry by adjusting either the initial electron spin polarization [Fig. \ref{fig. gae} (c)] or the ellipticity of the laser fields [Fig. \ref{fig. gae} (d)]. When the incident electrons are in a pure state with \(\zeta_y = -1\), most axions are deflected to \(\theta_{y,\max} = -0.58\) mrad [Fig. \ref{fig. agl} (b)]. As the average polarization degree \(|\overline{\zeta}_y|\) decreases by introducing electrons with \(\zeta_y = 1\), the number of axions deflected to \(\theta_y > 0\) increases, leading to a reduction in the asymmetry \(|R|\) [Fig. \ref{fig. gae} (c)]. However, the central deflection angle \(\theta_{y,\max}\) of the density peak remains unchanged until \(\overline{\zeta}_y\) becomes positive. 
In addition, as the laser ellipticity increases from $\epsilon=-0.2$ to 0.2, the central deflection angle increases monotonically from negative to positive values, owing to the dependence of the transverse momentum \( p_y \) on the field strength \( E_y \).
The asymmetry \( R \) vanishes when the laser is linearly polarized but starts increasing rapidly as ellipticity is introduced  [Fig. \ref{fig. gae} (d)].  With the increase of ellipticity $\epsilon$, the impact of $E_y$ on electron's motion becomes dominate over other effects (e.g.  pondermotive force and radiation reaction), leading to a more notable correlation of $p_y$ and $B_y$ and consequently larger $R$. As \(\epsilon \) continues to increase, the growth of \( R \) slows down, eventually reaching saturation for \( \epsilon \geq 0.1 \). 
Thus, tuning the laser ellipticity allows precise control over the central peak of the deflection angle, whereas the density asymmetry is more strongly influenced by the electron polarization.

\begin{figure}[b]
    \centering
    \includegraphics[width=0.5\textwidth]{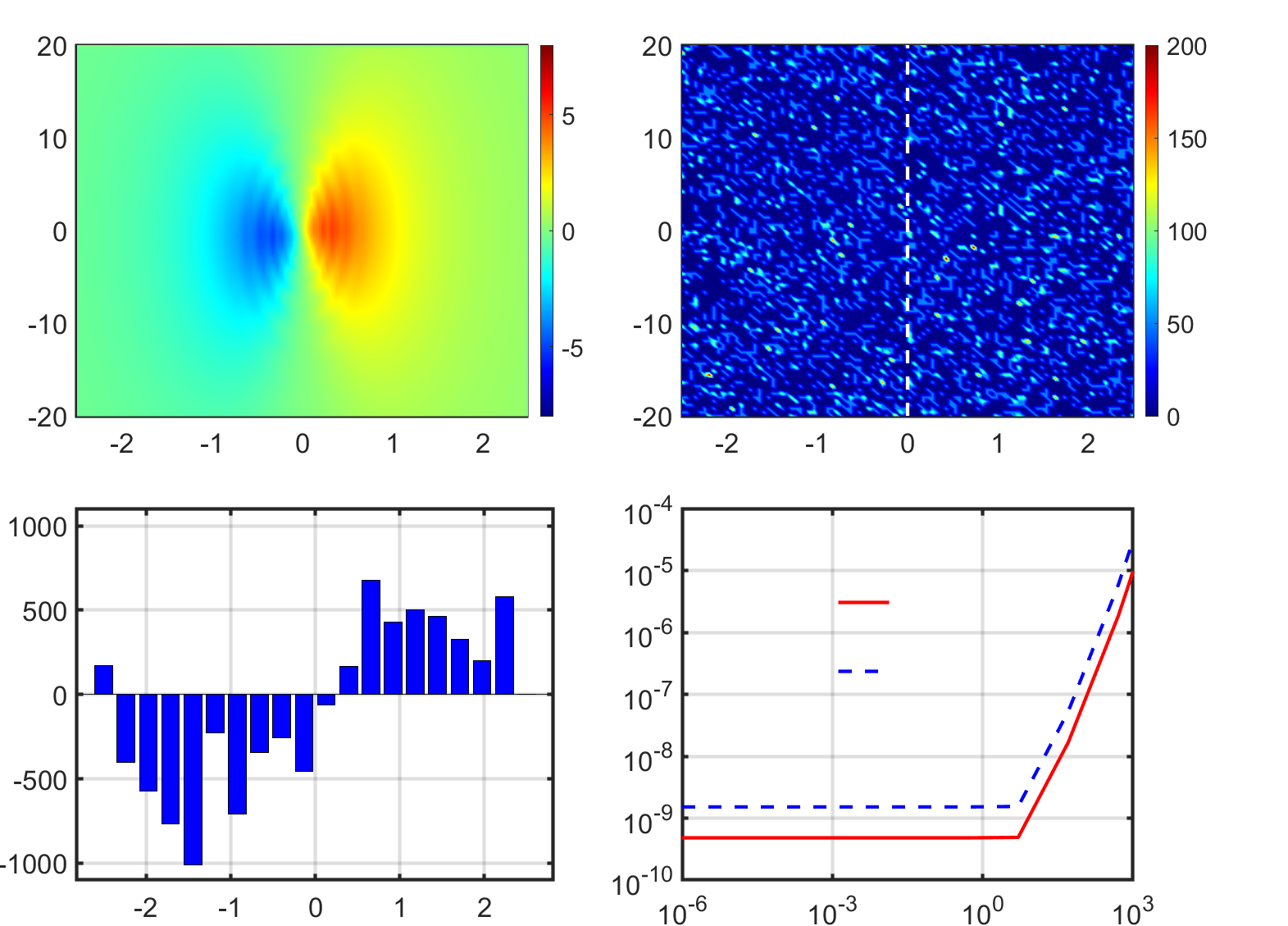} 
  \begin{picture}(300,20)
    \put(20,185){(a)}
    \put(137,185){\textcolor{white}{(b)}}
    \put(20,95){(c)}
    \put(137,95){(d)}
    \put(-5,140){\rotatebox{90}{$\theta_x$ (mrad)}}
    \put(45,107){$\theta_y$(mrad)}

    \put(165,107){$\theta_y$(mrad)}
    \put(191,130){\textcolor{white}{$\zeta_{y}=1$}}
    \put(143,130){\textcolor{white}{$\zeta_{y}=-1$}}
    \put(-8,57){\rotatebox{90}{$\Delta N$}} 
    \put(45,15){$\theta_y$ (mrad)}
    \put(113,35){\rotatebox{90}{$g_{ae}g_{a\gamma\gamma}$ (GeV$^{-1}$)}}
    \put(167,14){$m_{a}$(eV)}
    \put(177,83){\small{10 years}}
    \put(177,69){\small{1 year}}

    \end{picture}
    \caption{(a) Difference of the regenerated photon yield defined as \(\Delta N = d^2 N_\gamma(\zeta_y = +1)/d\theta_x d\theta_y - d^2N_\gamma(\zeta_y = -1)/d\theta_x d\theta_y\), versus \(\theta_x\) and \(\theta_y\). 
(b)Angular distribution \( d^2N_\gamma / d\theta_x d\theta_y \) (mrad\(^{-2}\)) including background noise, shown for electron spin states \(\zeta_y = -1\) (left) and \(\zeta_y = +1\) (right).
(c) Asymmetry \(\Delta N\) derived from the difference between the right (\(\zeta_y = +1\)) and left (\(\zeta_y = -1\)) halves of panel (b).  $g_{ae}g_{a\gamma\gamma}(\text{GeV}^{-1})=1.45\times 10^{-8}$ for (a)-(c).
(d) Projected exclusion limit  at the 95\% confidence level on the coupling \( g_{ae}g_{a\gamma\gamma} (\text{GeV}^{-1}) \) as a function of the axion mass \( m_a \), assuming a measurement time of one year (blue dashed line) and ten years (red solid line). 
}
    \label{fig. bound} 
\end{figure}
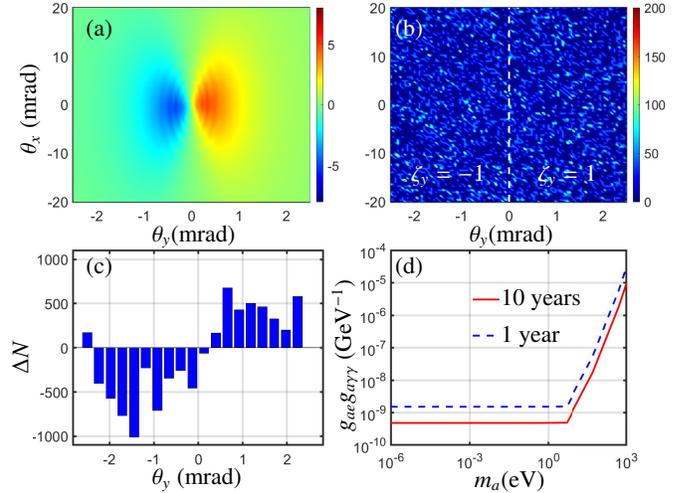

In our scheme, copious $\gamma$ photons are generated due to the electron-photon coupling, which can also contribute to the axion generation via the Primakoff effect. To include this process, we derived the conversion rate using perturbertive QED calculations \cite{SM3}, and 
 evaluated it through integration over the trajectories of the emitted $\gamma$ photons.
In the ultrarelativistic limit, the axion's momentum direction approximately aligns with that of the $\gamma$ photon. 
Since the spin-dependent asymmetry in photon emission is much weaker than in axion generation, the angular distribution of $\gamma$ photons is nearly symmetric about $\theta_y=0$; see Fig. \ref{Fig.garr} (a). 
Consequently, the angular distribution of axions generated by the axion-photon coupling is also insensitive to the spin state of electrons; see Fig. \ref{Fig.garr} (d).  Meanwhile, \(\gamma\) photons from nonlinear Compton scattering are characterized by substantial soft photon emission [Fig. \ref{Fig.garr} (b)]. Unfortunately, the conversion efficiency of these soft photons are rather low due to the suppressed  conversion probability and vanishing resonant bandwidth $\Delta \omega$ for small $m_a$ \cite{beyer2020axion}. Therefore, the generated axion $N_a\propto N_\gamma P_{a\gamma\gamma}$ has a maximum for varying axion mass, where $N_\gamma$ is the photon number within the resonant range $\Delta \omega$, and $P_{a\gamma\gamma}$ conversion probability. The simulation results show that the conversion peak appears at  \(\delta_m = 0.03\); see Fig. \ref{Fig.garr} (c). The angular distribution of axion for \(\delta_m = 0.03\) are presented in Fig. \ref{Fig.garr} (d). The axion generated by axion-photon coupling is one orders smaller compared with that generated by axion-electron coupling. Therefore, the angular asymmetry induced by the spin effects survives even though Primakoff effect is taken in to account.

The produced axions can be detected by their reconversion into photons under strong magnetic fields. For a typical configuration with $B=5\text{T}$ and $L=4.21 \text{m}$, the angular distribution of the regenerated photons inherits the spin-induced asymmetry of the axions  [Fig. \ref{fig. bound} (a)]. 
In conventional axion-search experiments, the sensitivity is primarily limited by Poisson fluctuations in the background photon count rate (typically, $n_b=10^{-4}\text{s}^{-1}$ \cite{bahre2013any}) [Fig. 5(b)]. In contrast, our scheme reveals a dipole-like structure in the photon distribution by subtracting yields associated with opposite spin polarizations [Fig. 5(c)]. This spin-dependent asymmetry offers an additional and robust signature of axion signals over background noise. Moreover,  the strong laser field can produce a substantial single-shot yield of axions ($N_a\gtrsim 10^{10}g_{ae}^2$), with an ultrashort signal duration on the femtosecond scale, thereby yielding a high signal-to-noise ratio. Considering the repetition rate of a 10 PW laser facility being 1/60 Hz, and a measure time of 1 year, the projected sensitivity bounds for the coupling $g_{ae}g_{a\gamma\gamma}~ (\text{GeV}^{-1})$ is 
$\sim 10^{-9}$ [Fig. \ref{fig. bound} (d)]. 


In conclusion, we propose a spin-controlled scheme for axion generation via nonlinear Compton scattering in strong laser fields. By tuning the electron polarization or laser ellipticity, the axion emission angle becomes continuously adjustable, leading to an asymmetric angular distribution. This built-in directional asymmetry not only constitutes a distinctive experimental signature, but also acts as a self-referencing null test, enabling robust discrimination against isotropic noise, astrophysical $\gamma$-ray backgrounds, and conventional QED signals. 
Our proposal may contribute to laboratory-based constraints on the axion parameter space by providing improved control over axion generation  through spin effects and by enhancing the axion yield via access to the strongly nonlinear interaction regime. 
In the future, the sensitivity of this proposal can be systematically improved by increasing the laser repetition rate  and electron bunch density to generate more axions. More promisingly, by accessing larger values of $\chi_e$,  the axion yield from nonlinear Compton scattering could increase by several orders of magnitude compared to the currently investigated regime~\cite{he2025semiclassical}, and pair production processes may also be triggered, further boosting axion production and facilitating the experimental verification.




{\it Acknowledgement:}
This work is supported by the National Natural Science Foundation of China (Grants No.12474312), and the National Key R\&D Program of China (Grant No. 2021YFA1601700). P.-L. H. thanks the sponsorship from Yangyang Development Fund.
\vspace{10pt}

\bibliography{gea}

\end{document}